\documentclass[usenatbib]{mn2e}  %Manuscript ID: MN-08-0046-L sub 14/1/08 resub 7/2/08
\usepackage{natbib,psfig}
\def\fracj#1#2{{\textstyle{#1\over#2}}}
\catcode`\@=11
\def\gta{\ifmmode{\,\mathrel{\mathpalette\@versim>\,}}
    \else{$\,\mathrel{\mathpalette\@versim>}\,$}\fi}
\def\lta{\ifmmode{\,\mathrel{\mathpalette\@versim<\,}}
    \else{$\,\mathrel{\mathpalette\@versim<}\,$}\fi}
\def\@versim#1#2{\lower 2.9truept \vbox{\baselineskip 0pt \lineskip
    0.5truept \ialign{$\m@th#1\hfil##\hfil$\crcr#2\crcr\sim\crcr}}}
\catcode`\@=12  
\def\b#1{{\bf#1}}
\def\vlos{v_\parallel}
\def\vperp{v_\perp}
\def\Flos{F_\parallel}

\def\kms{\,{\rm km}\,{\rm s}^{-1}}

\def\d{{\rm d}}

\title[Fitting orbits to tidal streams]
{Fitting orbits to tidal streams}

\author[J. Binney]{James  Binney\\
Rudolf Peierls Centre for Theoretical Physics, Keble Road, Oxford OX1 3NP, UK\\}

\begin{document}

\date{Draft, February 7, 2008}

\pagerange{\pageref{firstpage}--\pageref{lastpage}} \pubyear{2007}

\maketitle

\label{firstpage}

\begin{abstract}
Recent years have seen the discovery of many tidal streams through the
Galaxy. Relatively straightforward observations of a stream allow one to
deduce three phase-space coordinates of an orbit. An algorithm is presented
that reconstructs the missing phase-space coordinates from these data.  The
reconstruction starts from assumed values of the Galactic potential and a
distance to one point on the orbit, but with noise-free data the condition
that energy be conserved on the orbit enables one to reject incorrect
assumptions. The performance of the algorithm is investigated when errors
are added to the input data that are comparable to those in published data
for the streams of Pal 5. It is found that the algorithm returns distances and
proper motions that are accurate to of order one percent, and enables one to
reject quite reasonable but incorrect trial potentials. In practical
applications it will be important to minimize errors in the input data, and
there is considerable scope for doing this.

\end{abstract}


\section{Introduction}

Mapping the matter distribution in the outer reaches of the Galaxy, which is
expected to be dark-matter dominated, is currently one of the central tasks
in astrophysics.  Recent surveys of the outer Galaxy have
revealed numerous enhancements in the density of stars that
have been securely identified as tidal streamers from Galactic satellites
\citep{Odenkirchen,Majewski04,Fieldstars}. Such structures effectively
delineate the orbit in the Galactic force-field of the parent body
\citep{JohnstonHB,Odenkirchen03}. Knowledge of the phase-space coordinates along
an individual orbit provides strong constraints on the force-field that
confines the orbit, and thus on the matter distribution that gives rise to the
force field. Consequently, in recent years considerable effort has been
devoted to the problem of determining the orbit of a satellite from
observations of its tidal stream \citep{Law05,Fellhauer07}.

Since tidal streams are generally discovered as enhancements in the density
of stars on the sky, the phase-space coordinates of the satellite orbit that are easiest
to determine are its Galactic coordinates $[l(u),b(u)]$, where $u$ is an
arbitrary parameter that increases along the orbit. With a moderate
observational follow-up effort, it is always possible to augment these data
with line-of-sight velocities $\vlos(u)$. The other three coordinates
(heliocentric distance $s$ and proper motion components) are much harder to
obtain, and are generally either unknown or measured with very low accuracy.
Given the incomplete nature of the data, the standard approach to orbit
fitting is to adopt a gravitational potential, and then to seek an orbit in
this potential that is consistent with the data \citep[e.g.][]{Fellhauer07}. The fitting procedure is
generally trial-and-error, and the final fit is often imperfect
\citep[e.g.][]{Fellhauer06}. 

Here we show that given the three coordinates $(l,b,\vlos)$
along a section of an orbit in a gravitational potential $\Phi_0(\b
r)$, one can uniquely determine what the remaining three phase-space
coordinates must be if the data are to be part of an orbit in a trial
potential $\Phi_{\rm t}(\b r)$. If $\Phi_{\rm t}=\Phi_0$, the recovered
phase-space coordinates are consistent with conservation of energy, while
when the trial and true potentials differ, energy conservation appears to be
violated. Therefore this procedure can be used to hunt for the Galaxy's
gravitational potential, and yields predictions for distances and proper
motions along a tidal stream that can be tested observationally.

\section{The algorithm}

We work in the inertial coordinate system in which the Galactic centre is at
rest. Consequently, line-of-sight velocities in this frame will differ from
measured heliocentric velocities by the projection of the Sun's velocity
(whch we assume known) onto the line of sight.  Let $\b s$ be the position
vector of the satellite with respect to an observer at the position of the
Sun, and let $\b F_{\rm t}(\b s)$ be the trial gravitational acceleration
($\b F_{\rm t}=-\nabla\Phi_{\rm t}$). Then if the trial potential is
correct, we have $\ddot\b s=\b F_{\rm t}$, and $\vlos=\hat\b s.\b
v=\dot s$ so $\dot\vlos=\Flos+\b v.\d\hat\b s/\d t$, where $\Flos$ is the
component of $\b F_{\rm t}$ along the line of sight.
 Also 
\begin{equation}
\b v={\d (s\hat\b s)\over\d t}=\dot s\hat\b s+s{\d\hat\b s\over\d t}
\end{equation}
 so 
\begin{equation}\label{eq:2}
\dot\vlos= \Flos+{v^2-\vlos^2\over s}= \Flos+{\vperp^2\over s}.
\end{equation}
 Moreover, the component of $\b v$ in the plane of the sky $\b v_\perp$ satisfies
 \begin{equation}\label{eq:3}
|\vperp|^2=s^2\dot u^2\equiv s^2(\dot b^2+\cos^2b\,\dot l^2),
\end{equation}
 where we have now fixed the meaning of the parameter $u$ to be the angular
distance along the trajectory.  Eliminating $|\vperp|^2$ between
equations (\ref{eq:2}) and (\ref{eq:3}), we obtain a quadratic equation for
$\d t$ that has solution
 \begin{equation}\label{basicdt}
\d t={1\over2\Flos}\left(\d\vlos\pm\sqrt{\d\vlos^2-4s\Flos\d u^2}\,\right).
\end{equation}
 Thus if we guess $s$ at some point on the trajectory, we can solve for $\d
t$ between observations, and calculate $E$. We get a new value of $E$ for
each pair of data points. We can also update $s$ from $\d s=\vlos\d t$.

To obtain a robust numerical scheme we divide $\d s=\vlos \d t$ and equation
(\ref{basicdt}) by the differential of $ u$ to obtain the set of coupled ordinary differential
equations
 \begin{eqnarray}\label{basicde}
{\d t\over\d u}&=&{1\over2\Flos}\left({\d\vlos\over\d u}\pm
\sqrt{\left({\d\vlos\over\d u}\right)^2-4s\Flos}\,\right)\nonumber\\
{\d s\over\d u}&=&\vlos{\d t\over\d u}.
\end{eqnarray}
 The sign ambiguity in the first equation is resolved such that $t$
increases along the trajectory.

\section{Tests}

The algorithm was tested by reconstructing orbits from pseudodata obtained
by projecting orbits in a model potential.  The tests showed that it is
important to determine accurately the distance $ u$ along the trajectory to
each data point. This can be determined in a two-step process: we first
obtain a crude estimate
 \begin{equation}
\lambda_i=\sum_{j<i}\sqrt{(b_{j+1}-b_j)^2
+\cos^2[\fracj12(b_{j+1}+b_j)](l_{j+1}-l_j)^2}.
\end{equation}
 Then we fit cubic splines in $\lambda$ to the values $b_i=b(\lambda_i)$ and
$l_i=l(\lambda_i)$, and use the resulting functions of $\lambda$ to evaluate
numerically the integrals
 \begin{equation}
 u_i=\int_0^{\lambda_i}\d\lambda\,\sqrt{{b'}^2+\cos^2(b){l'}^2},
\end{equation}
 where a prime denotes differentiation with respect to lambda. Finally we
fit cubic splines to the values $b_i=b( u_i)$, etc., and use these
functions of $ u$ and  the fourth-order
Runge-Kutta algorithm to solve equations (\ref{basicde}).
At each integration step we
re-evaluate $E=\frac12(\vlos^2+s^2\dot u^2)+\Phi_{\rm t}$.

\begin{figure}
\psfig{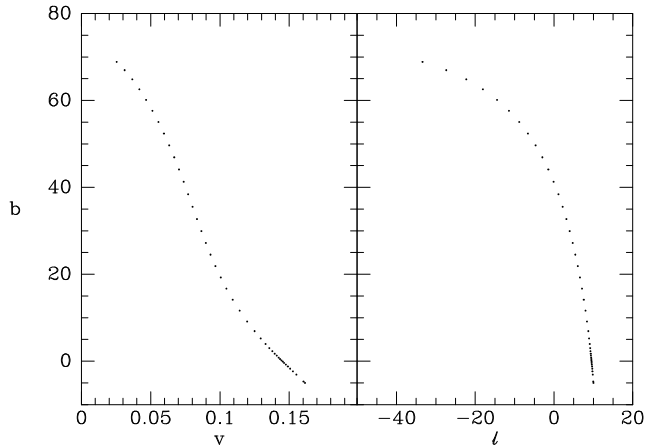}
\caption{Right panel: the projection onto the sky of a numerically integrated orbit in a
Miyamoto--Nagai potential with $b/a=0.2$. Left panel: the line-of-sight
velocity as a function of $b$.}
\label{onsky1}
\end{figure}

\begin{figure}
\psfig{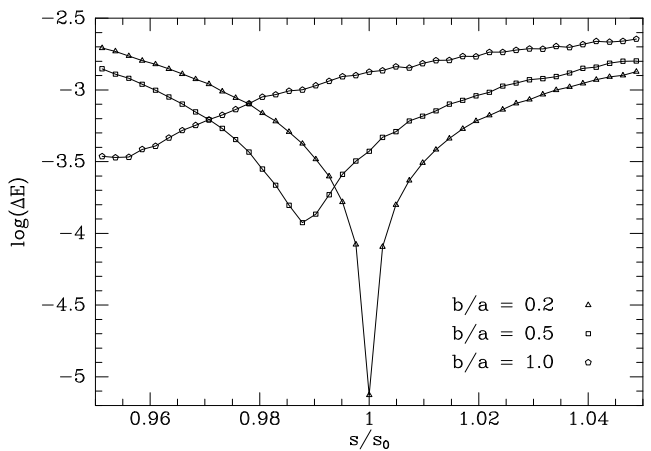}
 \caption{The log to base 10 of the rms variation in the energy when the
orbit shown in Fig.~\ref{onsky1} is reconstructed from an assumed initial
distance $s$ rather than its true value $s_0$. The triangles show results
obtained when the reconstruction employs true potential, while the squares
and pentagons are for less flattened potentials.}
\label{erms1}
\end{figure}

\begin{figure}
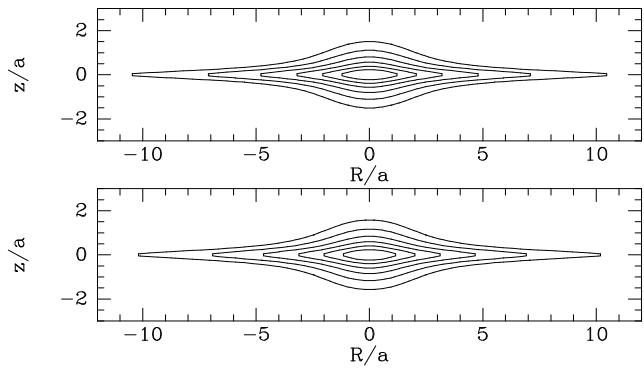

\psfig{file=miyam0p2.ps,width=\hsize}
\psfig{file=miyam0p22.ps,width=\hsize}
\caption{Isodensity contours in the meridional planes of Miyamoto--Nagai
models with $b/a=0.2$ (top) and $b/a=0.22$ (bottom). }
\label{Miyamfig}
\end{figure}

Orbits were integrated in the Miyamoto--Nagai potential that has $b/a=0.2$
and $GM/a=1$ \citep[and Fig.~\ref{Miyamfig}]{Miyam}.  They were then reconstructed using either
this potential or a Miyamoto--Nagai potential with different parameters.
The 42 dots in the right panel of Fig.~\ref{onsky1} show the input coordinates on the sky of
the model orbit when it is viewed from a point in the symmetry plane and
$8a$ from the centre, while the left panel shows the linbe-of-sight
velocities along the orbit. The orbit has energy $-0.0322GM/a$. It starts at
$b=-5\deg$ and a distance $15a$ from the point of observation, and moves out
to a distance $22.6a$ as it rises to $b=68.9\deg$.  The triangles in
Fig.~\ref{erms1} show the rms variation $\Delta E$ in the output energy $E$
along the reconstructed orbit when the true potential is used to integrate
equations (\ref{basicde}) from these data points and an initial distance $s$
that is the stated multiple of its true value, $s_0$.  There is a sharp
minimum in $\Delta E$ when $s=s_0$.  The squares and pentagons in
Fig.~\ref{erms1} show the corresponding results when the wrong potential is
used to reconstruct the orbit -- larger values of $b/a$ generate rounder
potentials.

When the potential is correct, we find to a good approximation that $\Delta
E\propto|1-s/s_0|$, while with an incorrect potential the curve of $\Delta
E$ can be fitted near its minimum with $\Delta E\propto\sqrt{(1-s/s_0)^2+k^2}$
with $k\ne0$. Thus in a practical case the quantity $k$ can be used as a
measure of how far the potential employed differs from its true value.

\begin{figure}
\psfig{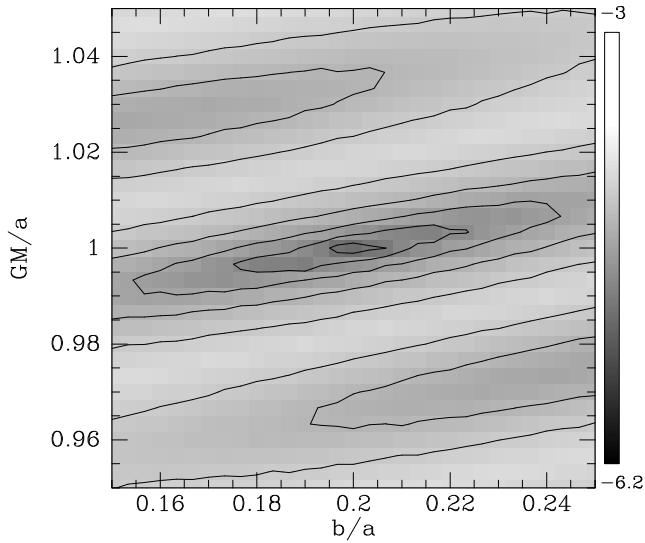}
\caption{A contour plot of $\log_{10}(\Delta E)$ as a function of the
parameters used in the reconstruction of the orbit segment shown in
Fig.~\ref{onsky1}.}
\label{erms2}
\end{figure}

Fig.~\ref{erms2} explores the effect of varying both the parameters $b/a$
and $GM/a$ of the potential used in the reconstruction of this orbit.
Although an error in one parameter can to some extent be offset by an error
in the other parameter, there is a sharp minimum in $\Delta E$ when the
parameters take their true values.  Thus from this one orbit one can
determine both the flattening and the mass of the potential with precision.
Fig.~\ref{Miyamfig} illustrates how high this precision is by showing
isodensity contours of the Miyamoto-Nagai models that have $b/a=0.2$ (top)
and $0.22$. Although these models are extremely similar, analysis of the
orbit segment easily distinguishes them.

Changing the length of the orbit segment analysed by a factor 3 either side
of the length shown in Fig.~\ref{onsky1} does not change the results
significantly. In particular, the algorithm identifies incorrect potentials
even with quite short orbit segments.

These tests demonstrate that with high-quality data the algorithm has no
difficulty recovering the true orbit if the potential is the correct one,
and provides excellent discrimination against incorrect potentials.

\subsection{Multiple orbit segments}
Different streams will be sensitive to different aspects of the Galactic
potential -- streams near the plane will be sensitive to the structure of
the disc, while the Magellanic Steam must be insensitive to the disc but
sensitive to the total Galactic mass. The Galactic potential would be most
strongly constrained by a plot like that shown in Fig.~\ref{erms2} in which
the quantity contoured is the sum of $\Delta E$ for all available streams.

Obscuration by dust near the plane sometimes breaks a stream into two or
more segments with an unobserved gap between them. In principle each segment
could be analysed independently, but this would not be the optimal
procedure.  First, all segments should be used to determine the path of the
underlying orbit on the sky, including through obscured regions. Then the
algorithm should be used to reconstruct each segment, and the rms energy
variation along all reconstructed segments (treated as a whole)
minimized within the space spanned by the initial distances assumed for each
section.

\subsection{Effect of errors in the data}

In practice the input data will be less exact than in the tests above
because tidal streams have finite widths on the sky and do not exactly trace
a single orbit. In favourable cases they delineate an orbit with great
precision: for example the tidal stream of a globular cluster such as that
of Pal 5 is precisely bounded by the cluster's orbit \citep{Odenkirchen03}. The
tidal streams of larger bodies, such as the Sgr dwarf galaxy, are broad, and
their relation to the orbit of the progenitor is less evident.
Correspondingly, in the case of a globular cluster, the uncertainty in the
radial velocity along the stream will be limited to measurement errors in
the velocities of stars, while in the case of a dwarf galaxy one has to
contend also with the velocity dispersion of the progenitor, which both
widens the range of velocities encountered at any point on the stream and
leads to a systematic offset between the mean velocity in the stream and the
velocity of the underlying orbit.

\begin{figure}
\psfig{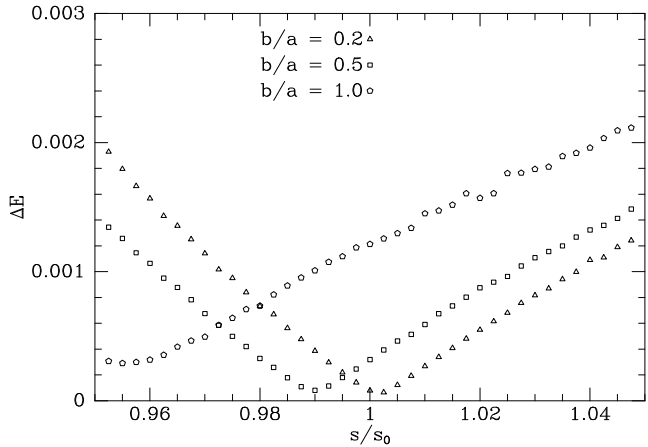}
\caption{As for Fig.~\ref{erms1} except that equation (\ref{erroreq}) has
been used to add errors to $b$ and $l\cos(b)$ that each amount to
$2\,$arcmin rms. The minimum value of $\Delta E$ for $b/a=0.5$ is bigger
than for $b/a=0.2$ by a factor $1.3$.}
\label{ermsdb}
\end{figure}

\begin{figure}
\psfig{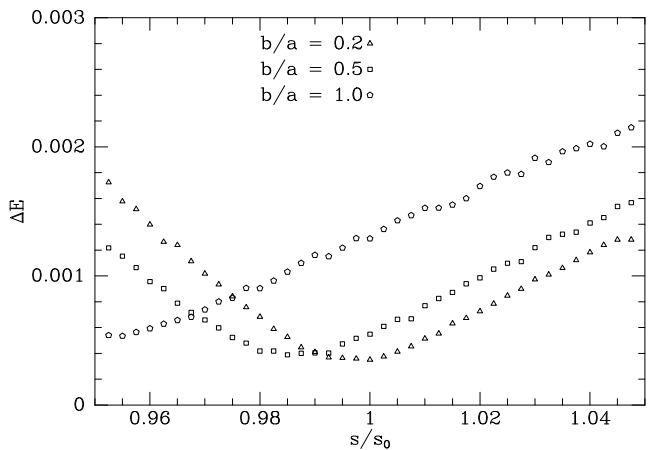}
\caption{As for Fig.~\ref{erms1} except that equation (\ref{erroreq}) has
been used to add errors to $\vlos$ that amount to 0.1 percent of the
circular speed at the start of the segment. The minimum value of $\Delta E$
for $b/a=0.5$ is bigger than for $b/a=0.2$ by a factor $1.1$.}
\label{ermsdv}
\end{figure}

When applying the algorithm to a stream such as that of Pal 5, one would not
use the positions and velocities of individual stars directly but would make
a judgment as to the relation of the orbit to the stellar stream, and
estimate the radial velocity on the orbit by averaging the velocities of
nearby stars and then making a correction for the expected offset between
the stream and orbital velocities. When all this had been done, one would
contrive that the trajectory of the orbit on the sky and in radial velocity
was smooth.  Therefore, in tests it is not appropriate to ad independent
errors to each data point, but rather to add correlated errors that leave
the trajectory through $(l,b,\vlos)$ space smooth. We do this as follows. For
$x=\vlos,b,l\cos b,$ we add a Fourier series 
 \begin{equation}\label{erroreq}
\delta x( u)=\sum_nX_n\cos(2n\pi\hat u+\xi_n),
\end{equation}
 where $n=0-9$ is an integer, $\hat u= u/ u_{\rm max}$ ranges from zero to
unity along the trajectory, and $\xi_n$ is randomly distributed in $(0,2\pi)$.
The amplitudes $X_n$ determine the power spectrum of the noise. We have no
way of knowing what this would be, but have assumed that
$X_n\propto(1+n)^{-1.1}$.

Studies of the tidal stream of Pal 5 provide a guide to the magnitude of the
errors to be encountered in real data. The stream's full width is
$\sim30\,$arcmin \citep{Odenkirchen03}, so it should be possible to
determine the position of the underlying orbit with an error that does not
exceed a tenth of this, and could be substantially smaller.
\cite{Odenkirchen02} report velocities along the stream that have errors
$\sim0.15\kms$ plus a dispersion of order $0.9\kms$ arising from binarity.
The latter dispersion could be beaten down by making observations at several
epochs; even at one epoch, measurements of many stars would enable the
velocity of the underlying orbit to be obtained to a precision of at least
$0.2\kms$. Hence we focus here on  errors of $2\,$arcmin in $b$ and
$l\cos(b)$, and errors in $\vlos$ that are $0.1$
percent of the circular speed at the start of the orbit segment.

\begin{figure}
\psfig{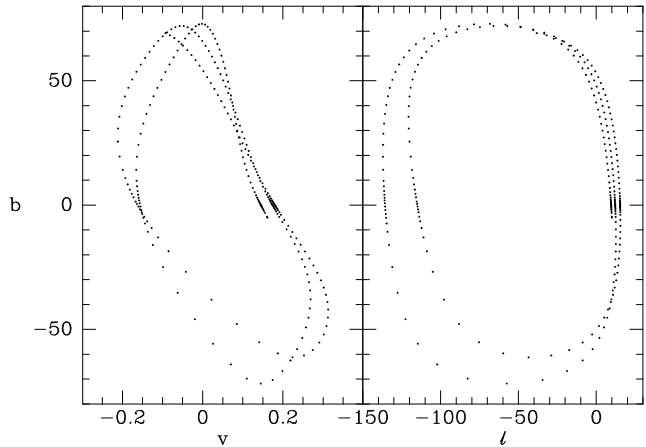}
\caption{Data points for an orbit segment ten times as long as in
Fig.~\ref{onsky1}.}
\label{onsky2}
\end{figure}

\begin{figure}
\psfig{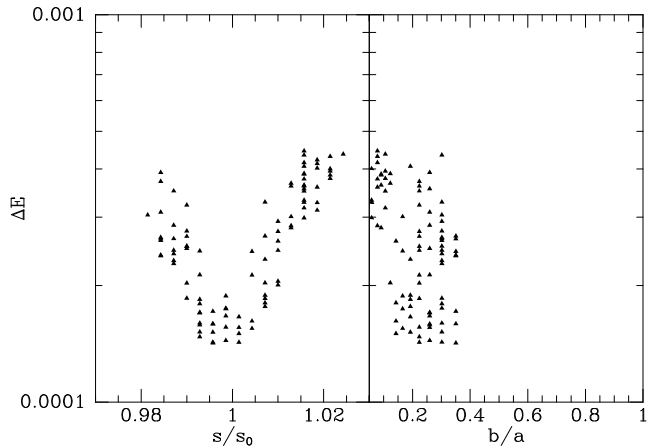}
\caption{The results of reconstructing 100 realizations of the orbit segment
of Fig.~\ref{onsky2}, with astrometric errors of 2 arcmin and velocity errors
of 0.1 percent of $v_{\rm c}$. The points show the values of $s/s_0$ and
$b/a$ that gave the smallest value of $\Delta E$, together with that value
of $\Delta E$.}
\label{rbscatter}
\end{figure}

Fig.~\ref{ermsdb} shows that with these errors in $b$ and $l\cos(b)$ it is
still possible to determine the distance to the orbit with exquisite
accuracy, and to tell that the potential with $b/a=0.2$ is to be preferred
to that with $b/a=0.5$.

\begin{figure}
\psfig{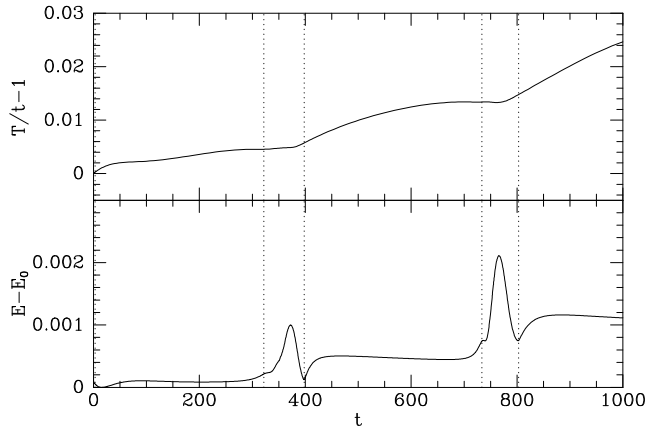}
\caption{Variation in  reconstructed energy $E$ and time $T$ along the orbit
shown in Fig.~\ref{onsky2} when the correct potential and distance are used
when $\vlos$ is in error by $0.1$ percent of $v_{\rm c}$. The vertical
dotted lines mark the times when the orbit passes through the plane.}
\label{dEdT}
\end{figure}

Fig.~\ref{ermsdv} shows the results obtained when the radial velocities are
in error by $0.1$ percent of the circular speed at the starting point of the
segment.  The impact of these velocity errors is larger than that of the
position errors. However, it is still just possible to reject the potential
with $b/a=0.5$, and the distance is accurately determined even if the wrong
potential is chosen.  When the orbit segment used for the reconstructions
covers a time interval that is longer by a factor ten (Fig.~\ref{onsky2}),
it becomes possible to reject the potential with $b/a=0.5$ with security.
Fig.~\ref{rbscatter} illustrates this finding by showing for each of 100
different realizations of the noise added to the positions and velocities of
Fig.~\ref{onsky2} the best fitting values of $s/s_0$ and $b/a$ together with
the associated values of $\Delta E$. All distance estimates lie within 2
percent of the truth; the mean of $s/s_0=1.002$ while its rms dispersion
is $ 0.012$. Values of $b/a$ in the range $(0.1,1)$ were searched but no
value bigger that $0.35$ was recovered; the mean of $b/a=0.201$ while
its rms dispersion was $0.095$. Thus with these errors in the data it is
possible to say that the potential is highly flattened, but not actually
determe $b/a$.

Fig.~\ref{dEdT} shows the reconstructed values of time and energy along the
best reconstruction of the long segment plotted in Fig.~\ref{onsky2} when
the velocities are in error by $0.1$ pecent of the circular speed.
Vertical dotted lines mark the instants at which the orbit passes through
the plane; pericentre and the point of closest approach to the Sun lie
between pairs of these plane-crossings. Some combination of plane crossing
and the peak in the proper motion as the orbit passes closest to the Sun
induces features in both the time and energy plots. These features are
independent of the accuracy parameter specified for the integration of
equations (\ref{basicde}), and they have no analogues when error-free data
are used for the reconstruction. The reconstructed time
$T$ is systematically later than it should be by a few percent, and
the reconstructed energy
is correspondingly larger than it should be.

Since the proper motions that the algorithm predicts are inversely
proportional to the elapse of $T$ along the stream, the figure shows that
even with errors in the input velocities, proper motions are generally
accurate to a few percent.

The algorithm's sensitivity to errors in $\vlos$ arises because the
underlying physical idea is that the change in $\vlos$ between data points
requires a certain time interval, and this time interval determines both the
change in distance between the points and the orbit's proper motion. Thus
the reconstruction is driven by differences in $\vlos$, which are sensitive
to errors in $\vlos$. When equation (\ref{erroreq}) is used to simulate
errors, the error in the difference in $\vlos$ between neighbouring points
scales like $ u_{\rm max}^{-1}$, so longer orbit segments produce more
accurate results.

\section{Conclusions}

A tidal stream delineates the orbit of its progenitor, so it can be used to
obtain three phase-space coordinates of an orbit. To complete the
specification of the orbit, one needs to associate with each of its points
on the sky a distance and a time of passage. The algorithm presented here
yields these additional quantities for an assumed potential and a distance
to one point on the stream. Unless these two inputs are correct, there is no
guarantee that energy will be conserved along the reconstructed orbit, and
indeed numerical experiments show that with noise-free data, the rms
variation in energy along the reconstructed orbit allows one to identify the
potential and determine the correct initial distance.

The algorithm is less sensitive to realistic errors in the
position of the stream on the sky than to likely errors in
$\vlos$ because the reconstruction is driven by {\it differences\/} in
$\vlos$ between grid points. 

The algorithm predicts the distances and proper motions of objects in the
stream, so when these data are available, they will provide an interesting
check on the correctness of all inputs and assumptions. Crude distance
estimates are generally available and seriously erroneous potentials can
probably be excluded because they return astrophysically unacceptable
distances. With presently attainable accuracy of the input
data it should be possible to determine distances to with about one percent,
and to constrain the Galactic potential quite strongly.

It is important to recognize that the inputs to the algorithm should be the
sky coordinates and line-of-sight velocities of the orbit of a test
particle, not the corresponding coordinates of a stream; streams delineate
orbits, but they do not actually lie along them. In practice a significant
effort will be required to extract the optimum input data from measurements
of any given stream, and an optimal approach to this task does not appear to
have been published.
However, the experiments presented here indicate that the rewards
for obtaining accurate input data will be very high: even one orbit segment
provides a wealth of precision information about the Galactic potential and
observationally testable predictions about the stream. The Galactic
potential could be even more strongly constrained by analyzing  several orbit
segments together.

\section*{Acknowledgments}
It is a pleasure to thank Andy Eyre, John Magorrian and the referee, Konrad
Kuijken,  for valuable
suggestions.

\label{lastpage}

\end{document}